\documentclass[preprint,showpacs,preprintnumbers,amsmath,amssymb]{revtex4}

\newcommand{\ds}{\displaystyle}

\usepackage{graphicx}
\usepackage{dcolumn}
\usepackage{bm}

\newcommand{\be}{\begin{equation}}
\newcommand{\en}{\end{equation}}
\newcommand{\bea}{\begin{eqnarray}}
\newcommand{\ena}{\end{eqnarray}}

\begin{document}


\title{On the Stability of Jordan-Brans-Dicke Static Universe}

\author{Sergio del Campo}
 \email{sdelcamp@ucv.cl}
\affiliation{ Instituto de F\'{\i}sica, Pontificia Universidad
Cat\'{o}lica de Valpara\'{\i}so, Casilla 4059, Valpara\'{\i}so,
Chile.}
\author{Ram\'on Herrera}
\email{ramon.herrera@ucv.cl} \affiliation{ Instituto de
F\'{\i}sica, Pontificia Universidad Cat\'{o}lica de
Valpara\'{\i}so, Casilla 4059, Valpara\'{\i}so, Chile.}
\author{Pedro Labra\~{n}a}
 \email{plabrana@ubiobio.cl}
\affiliation{ Departamento de F\'{\i}sica, Universidad del
B\'{\i}o-B\'{\i}o, Avenida Collao 1202, Casilla 5-C, Concepci\'on,
Chile.}

\date{\today}

\begin{abstract}

In this work we study the stability of the Jordan-Brans-Dicke
(JBD) static universe. This is motivated by the possibility that
the universe might have started out in an asymptotically JBD
static state, in the context of the so called emergent universe
scenario.
We extent our previous results on stability of JBD static universe
by considering spatially homogeneous Bianchi type IX anisotropic
perturbation modes and by including more general perfect fluids.
Contrary to general relativity, we have found that the JBD static
universe, dominated by a standard perfect fluid, could be stable
against isotropic and anisotropic perturbations. The implications
of these results for the initial state of the universe and its
pre-inflationary evolution are discussed.

\end{abstract}

\maketitle

\section{Introduction}

Measures coming from three different independent sources
(\mbox{CMB}, Type Ia supernova, and cluster abundances) strongly
suggest that the expansion of the universe has been accelerating
during recent epoch \cite{6}. This discovery has become one of the
main challenges for modern theoretical physics.

In general, dark energy or quintessence, responsible for the
cosmic acceleration, determines the features of the evolution of
the universe. Certainly, the nature of this sort of energy may
lead to the improvement of our picture about particle physics
and/or gravitation. Most of these studies have been done in the
standard theory of gravity, i.e. general relativity theory (GR).
However, motivated mainly from string theories, a less standard
theory have been carried out, namely the so called scalar-tensor
theory of gravity\cite{Jbd, stensor, stensor-fantasma, 48}.
An important advantage of these models is that they naturally
allow \cite{48,stensor-fantasma} a super-accelerating expansion of
the universe where the effective dark energy equation of state
$w=\frac{p}{\rho}$ crosses the phantom divide line $w=-1$. Such a
crossing is consistent with current cosmological data \cite{fant}
and if it is confirmed, it would become an enigmatic problem which
can not been explained easily with standard {\it quintessence}
models \cite{Gannouji:2006jm, Andrianov:2005tm}.

The archetypical theory associated with scalar tensor models is
the JBD gravity. The JBD theory\cite{Jbd} is a class of models in
which the effective gravitational coupling evolves with time. The
strength of this coupling is determined by a scalar field, the
so-called JBD field, which tends to the value $G^{-1}$, the
inverse of the Newton's constant. The origin of JBD theory is in
Mach's principle according to which the property of inertia of
material bodies arises from their interactions with the matter
distributed in the universe. In modern context, JBD theory appears
naturally in supergravity models, Kaluza-Klein theories and in all
known effective string actions \cite{Freund:1982pg,
Appelquist:1987nr, Fradkin:1984pq, Fradkin:1985ys, Callan:1985ia,
Callan:1986jb, Green:1987sp}.


The study of static universe and its stability has always been of
great interest since the pioneer work of Eddington
\cite{Eddington}.
For example, in the context of GR the stability of the Einstein
static (ES) universe in the presence of conventional matter field
has been studied in Refs.~\cite{Harrison:1967zz, Gibbons:1987jt,
Barrow:2003ni}. In the presence of ghost scalar field it was
studied in Ref.~\cite{Barrow:2009sj}.
Also, the ES universe has been studyied in different gravitational
theories. In GR a generalization which include a variable pressure
have been analyzed in Ref. \cite{Boehmer:2003uz}. In the context
of brane world models it was considered in Refs.
\cite{Gergely:2001tn, Lidsey:2006md, Banerjee:2007qi,
Hamilton:2008ey}. The study in the Einstein-Cartan theory it is
found in Ref. \cite{Boehmer:2003iv}. In loop quantum cosmology,
this subject has been studied in Refs. \cite{Mulryne:2005ef,
Parisi:2007kv, Nunes:2005ra}.
The stability of the ES universe in a $f(R)$ gravity and in
modified Gauss-Bonnet gravity theories have been studied in~Refs.
\cite{Boehmer:2007tr} and \cite{Boehmer:2009fc}, respectively.

Recently, the stability of ES models has become relevant for the
study of cosmological scenarios in which the ES universe
corresponds to an initial state for a past-eternal inflationary
cosmology, the so-called emergent universe scenario
\cite{Ellis:2002we}.
The original idea of an emergent
universe"~\cite{Ellis:2002we,Ellis:2003qz}
is that in which the universe emerges from an ES universe state
with radius $a_0 >> l_p$ (where $a_0$ is the scale factor at some
instant and $l_p$ is the Planck length), inflates and  then is
subsumed into a hot Big Bang era. Such models are appealing since
they provide specific examples of non–singular (geodesically
complete) inflationary universes. Also, these models could avoid
an initial quantum-gravity stage if the static radius is larger
than the Planck length.

However, the emergent universe models based on GR, with ordinary
matter, suffer from a number of important shortcomings. In
particular, the instability of the ES
state~\cite{Mulryne:2005ef,Banerjee:2007qi,Nunes:2005ra,Lidsey:2006md}
makes it extremely difficult to maintain such a state for an
infinitely long time. The instability of the ES solution ensures
that any perturbations, no matter how small, rapidly force the
universe away from the static state, thereby aborting the
scenario.

Some models have been proposed to solve the stability problem of
the asymptotic static solution. They consider non-perturbative
quantum corrections of the Einstein field equations, either coming
from a semiclassical state in the framework of loop quantum
gravity (LQG) \cite{Mulryne:2005ef,Nunes:2005ra} or braneworld
cosmology with a timelike extra dimension \cite{Lidsey:2006md,
Banerjee:2007qi}. Other possibilities to consider are the
Starobinsky model or exotic matter \cite{Mukherjee:2005zt,
Mukherjee:2006ds}.

On the other hand, it has been shown that a scalar tensor theory
could solve the problem of the instability of the emergent
universe models. In particular, in Ref.~\cite{delCampo:2007mp}, it
was found that a self interacting JBD theory presents a stable
past eternal static solution, which eventually enters a phase
where the stability of this solution is broken leading to an
inflationary period, providing in this way, an explicit
construction of an emergent universe scenario.

In this work, we study the stability of the JBD static universe.
This is motivated by the possibility that the universe might have
started out in an asymptotically JBD static state
\cite{delCampo:2007mp}.
We extent our previous results on the stability of JBD static
universe by consider spatially homogeneous Bianchi type IX
anisotropic perturbation modes and by including  more general
perfect fluids.
General anisotropic perturbations are important to be consider
because they could be the crucial destabilizing modes of a static
universe, see Refs. \cite{Gibbons:1987jt, Barrow:2003ni}.
Contrary to the GR case we have found that the JBD static universe
dominated by a perfect fluid could be stable against isotropic and
anisotropic perturbations for some sort of matter components.


The paper is organized as follows. In Sect.~\ref{secti} we review
briefly the cosmological equations of the JBD model. In
Sect.~\ref{static1} the existence and nature of static solutions
are discussed for universes dominated by a standard perfect fluid
and by a scalar field.
In Sect.~\ref{secanis} we study the stability of the JBD static
universe against small anisotropic perturbations. In
Sect.~\ref{secdin} we focus in a particular example of a JBD
potential which allow us to use a dynamical system approach to
study the problem of stability of the JBD static universe. In
Sect.~\ref{conc} we summarize our results.

\section{The Model \label{secti}}

We consider the following JBD action  for a self-interacting
potential and matter, given by\cite{Jbd}

 \begin{eqnarray}
 \label{ac1}
\ds S =  \int{d^{4}x\,\sqrt{-g}}\bigg[\frac{1}{2}\,\,\Phi\,R\,
-\,\frac{1}{2}\,\frac{w}{\Phi}\,\nabla_\mu\Phi \,\nabla^\mu\Phi
+\, V(\Phi) +{\cal{L}}_m\bigg],
 \end{eqnarray}
 where ${\cal{L}}_m$ denote the Lagrangian density of the matter,
$R$ is the Ricci scalar curvature, $\Phi$ is the JBD scalar field,
$w$ is the JBD parameter and $V(\Phi)=V$ is the potential
associated to the  field $\Phi$. In this theory $1/\Phi$ plays the
role of the gravitational constant, which changes with time. This
action also matches the low energy string action for $w=-1$
\cite{Green:1987sp}.

From the Lagrangian density, Eq. (\ref{ac1}), we obtain the field
equations:


\begin{eqnarray}
\label{fequation}
R_{\mu \nu} - \frac{1}{2}R\,g_{\mu \nu}\, -
\frac{w}{\Phi^2}\,\nabla_\mu\Phi\,\nabla_\nu\Phi &-&
\frac{1}{\Phi}\,\nabla_\mu\,\nabla_\nu \Phi + g_{\mu
\nu}\left(\frac{\square \Phi}{\Phi} +
\frac{w}{2\Phi^2}\left(\nabla \Phi \right)^2 -
\frac{V(\Phi)}{\Phi}\right) \nonumber\\ &=& \frac{1}{\Phi}\,T_{\mu
\nu}\,\,,
\end{eqnarray}

and

\begin{equation}\label{f2equation}
\Box \Phi = \frac{1}{3+2w}\,{T^\mu}_\mu + \frac{2}{3+2w}\,\left[2V
- \Phi \,\frac{dV}{d\Phi}\right],
\end{equation}

\noindent where we have consider that $\Phi$ is a function of the
cosmological time,$t$, only. Units are such that $c=\hbar=1$.


\section{The Static Universe Solution in JBD Theory\label{static1}}


Let us start by considering  the closed Friedmann-Robertson-Walker
metric:

 \be \ds
d{s}^{2}\,=\, d{t}^{2}\,-\, a(t)^{2}\left[\frac{dr^2}{1-r^2} +
r^2\,(d\theta^2 + \sin^2 \!\!\theta\,\, d\phi^2 )\right],
\label{met1} \en
where $a(t)$ is the scale factor, $t$ represents the cosmic time.
The matter content of the universe is modelled by a perfect fluid
with effective equation of state given by $P=(\gamma -1)\,\rho$.
In general, when the perfect fluid is described by a scalar field,
it is found that the parameter $\gamma$ becomes variable.
Thus, by using the metric, Eq.~(\ref{met1}), the set of field
equations~(\ref{fequation}) and (\ref{f2equation}) become
\begin{equation}
 H^2\,+\frac{1}{a^2}\,+H\frac{\dot{\Phi}}{\Phi}=\frac{\rho}{3\,\Phi}+
 \frac{w}{6}\left(\frac{\dot{\Phi}}{\Phi}\right)^2+\frac{V}{3\,\Phi}\label{key02},
\end{equation}

\begin{equation}
2\frac{\ddot{a}}{a}+H^2\,+\frac{1}{a^2}\,+\frac{\ddot{\Phi}}{\Phi}+2\,H\,\frac{\dot{\Phi}}{\Phi}+
 \frac{w}{2}\left(\frac{\dot{\Phi}}{\Phi}\right)^2-\frac{V}{\Phi} =-\frac{P}{\Phi}\label{dda},
\end{equation}
and
\begin{equation}
\ddot{\Phi}+3H\dot{\Phi}=\frac{(\rho-3
P)}{(2w+3)}+\,\frac{2}{2w+3}\left[2\,V-\Phi\,V'\right]\;.
\label{key_01}
\end{equation}
The energy-momentum conservation implies that
\begin{equation}
\dot{\rho}+3H(\rho+P)=0 \,,
\end{equation}
where $V'=dV(\Phi)/d\Phi$. Dots mean derivatives with respect to
the cosmological time.

In the context of JBD theory the static solutions  are closed
universes characterized by the conditions $a=a_0=Const.$,
$\dot{a}_0=0=\ddot{a}_0$ and $\Phi=\Phi_0=Cte.$,
$\dot{\Phi}_0=0=\ddot{\Phi}_0$, see Ref. \cite{delCampo:2007mp}.

Then the  static solution for a universe dominated by a general
perfect fluid is obtained if the following conditions are
fulfilled

\begin{eqnarray}
\label{v2} a_0^2 &=& \frac{3}{V'_{0}}\,, \\
\nonumber \\
\label{v} \rho_0 &=& V'_{0}\,\Phi_0 -\,V_0\,,\end{eqnarray} and
\begin{eqnarray}\label{v1g} \gamma_0 &=&
\frac{2}{3}\left(1+\frac{V_0}{\rho_0}\right)
2\,\frac{\Phi_0}{a_0^2\,\rho_0}\,,
\end{eqnarray}
where  $V_0=V(\Phi_0)$ and
\mbox{$V_0'=(dV(\Phi)/d\Phi)_{\Phi=\Phi_0}$}. These equations
connect the equilibrium values of the scale factor and the JBD
field with the energy density and the JBD potential at the
equilibrium point.

Note that in order to obtain a static solution we need to have a
non-zero JBD potential with a non-vanishing derivative at the
static point $\Phi=\Phi_0$. The original Brans-Dicke model
corresponds to $V(\Phi)=0$. However, non-zero $V(\Phi)$ is better
motivated and appears in many particle physics models. In
particular, $V(\Phi)$ can be chosen in such a way that $\Phi$ is
forced to settle down to a non-zero expectation value, $\Phi
\rightarrow m_p^2/8\pi$, where $m_p = 10^{19} GeV$ is the present
value of the Planck mass. On the other hand, if $V(\Phi)$ fixes
the field $\Phi$ to a non-zero value, then time-delay experiments
place no constraints on the Brans-Dicke parameter $w$
\cite{La:1989pn}.
In particular, if we choose the JBD potential in such a way that
$\Phi$ will be stabilized at a constant value, let say $\Phi_f$,
at the end of the inflationary period (see Ref.
\cite{delCampo:2007mp} as an  example), we can recover GR by
setting $\Phi_f = m_p^2/8\pi$, together with an appropriated value
for the parameter $w$ which will be in agreement with the solar
system bound \cite{La:1989pn, Sen:2001ki}.


\subsection{JBD static universe dominated by a standard perfect fluid}


The static solution is characterized by the Eqs.
(\ref{v2}-\ref{v1g}), from which we obtain $\gamma \geq
\frac{2}{3}$ as a condition for a static solution,
if the JBD potential, $V(\Phi)$, is
positive\cite{delCampo:2007mp}. Notice that this means that it is
not possible to have a static solution if the universe is
dominated by the cosmological constant (corresponding to $\gamma
=0$ in the equation of state), but it is possible to have a static
universe when it is dominated by dust or radiation, among others
possibilities.

Now, we study the stability of this solution against small
homogeneous and isotropic perturbations. In order to do this, we
consider small perturbations around the static solution for the
scale factor and the JBD field. We set

\begin{equation} \label{s1b}
a(t)=a_0\left[1+\varepsilon(t)\right],
\end{equation}
and
\begin{equation} \label{s2b}
\Phi(t)=\Phi_0\,[1+\beta(t))]\,.
\end{equation}
Then, we have
\begin{equation} \label{p1b}
\rho=\rho_0+\delta\rho(\varepsilon)\approx\rho_0-3\gamma\,\rho_0\,\varepsilon
 \,,
\end{equation}
where $\varepsilon\ll 1$ and $\beta\ll 1$ are small perturbations.
By introducing expressions (\ref{s1b}), (\ref{s2b}) and
(\ref{p1b}) into Eq.~(\ref{dda}) and Eq.~(\ref{key_01}), and
retaining terms at the linear order on $\epsilon$ and $\beta$, we
obtain the following coupled equations

\begin{equation} \label{s3b}
\ddot{\varepsilon} - \left[ \frac{1}{a_0^2} + 3\,\frac{(\gamma
-1)}{a^2_0}\right]\varepsilon -
\frac{\ddot{\beta}}{2}-\frac{\beta}{a_0^2}=0\,,
\end{equation}
and
\begin{equation} \label{s4b}
(3+2w)\,\ddot{\beta}-\left(\frac{6}{a_0^2}-2\,\Phi_0\,V_0''\right)\beta
+ (4 - 3\,\gamma)\,\frac{6}{a_0^2}\, \varepsilon=0\,,
\end{equation}
where $V_0''=(d^2V(\Phi)/d\Phi^2)_{\Phi=\Phi_0}$.

From the system of Eqs.(\ref{s3b}) and (\ref{s4b}) we can obtain
the frequencies for small oscillations


\begin{eqnarray}\label{pc2b}
\omega_{\pm}^2 &=& \frac{1}{a_0^2(3+2w)}\bigg[a_0^2\Phi_0\,V_0''
-6 + w\,(2 -
3\,\gamma) \\\nonumber \\
&\pm& \sqrt{\big[-6 + a_0^2\Phi_0V_0'' + 2w - 3w\,\gamma \big]^2 +
2(3+2w)\,\big(-6 + a_0^2\Phi_0V_0''\,\big[3\gamma
-2\big]\big)}\bigg].\nonumber
\end{eqnarray}

\noindent Note that the static solution is stable if the
inequality, $\omega_{\pm}^2>0$, is fulfilled. Assuming that the
parameter $w$ satisfies the constraint, $(3+2w)>0$, it is found
that the following inequalities must be achieved in order to have
a stable static solution

\begin{equation}\label{cond21}
\frac{2}{3}<\gamma <\frac{4}{3}\,, \,\,or\;\,\frac{4}{3}<\gamma\,,
\end{equation}

\begin{equation}\label{cond22}
-\frac{3}{2} < w < -18\,\frac{(\gamma-1)}{(2-3\gamma)^2}\,,
\end{equation}
and
\begin{equation}\label{cond23}
2(6+w)-3(3+w)\gamma + \sqrt{3}\,\big|4-3\gamma\big|\sqrt{3+2w} \,<
a^2_0\,\Phi_0\,V_0'' \,< \frac{6}{3\gamma-2}\,.
\end{equation}
From these inequalities we can conclude that for a universe
dominated by a standard perfect fluid (with $\gamma > 2/3$), it is
possible to find a solution where the universe is static and
stable.

Here, the only exception is radiation, where $\gamma=4/3$, which
becomes explicitly excluded by the latter inequalities.
This peculiar behavior for radiation could be understood due to
the particular way in which the perfect fluid appears in the
equation for the JBD field, Eq.~(\ref{key_01}), where it becomes
independent of the energy density and pressure of the fluid.
Then, we can note that in the radiation case $w_\pm^2$ are both
real numbers. On the other hand, $w_+^2$ could be a positive
number, but $w^2_-$ is always negative. Therefore, in this case,
we have a saddle instability.

\subsection{Scalar fields}

In the context of emergent universe models, the static JBD
universe dominated by a scalar field (inflaton) was studied in
Ref.~\cite{delCampo:2007mp}. Here, we reproduce the main results
concerning this static solution.

The energy density, $\rho$, and the pressure, $P$, are expressed
by the following equations
\begin{equation}
\rho = \frac{\dot{\Psi}^2}{2}+U(\Psi),
\end{equation}
and
\begin{equation} P = \frac{\dot{\Psi}^2}{2}-U(\Psi)\,.
\end{equation}
Here, $U(\Psi)$ represents the scalar potential associated to the
scalar field $\Psi$.

We could write an effective equation of state for the scalar
field, $\Psi$, expressed by the equation $P=(\gamma -1)\,\rho$,
where the equation of state "parameter", $\gamma$, could be
written as

\begin{equation} \label{gamma}
\gamma=2\left(1-\frac{U(\Psi)}{\rho}\right).
\end{equation}

During the static regimen, in the context of an emergent universe
models, the matter potential $U(\Psi)$ is consider as a flat
potential, that is $U(\Psi)=U_0=Const.$ and the scalar field rolls
along this potential with a constant velocity $\dot{\Psi}_0$.
The conditions for static universes, Eqs.~(\ref{v2}-\ref{v1g}),
imply that the following condition for the state parameter
\begin{equation}
\label{v1gb} \gamma_0 = 2\,\frac{\Phi_0}{a_0^2\,\rho_0}
2\left(1-\frac{U_0}{\rho_0}\right),
\end{equation}
must be satisfied.%

The velocity when the scalar field $\Psi$ is rolling along a
constant potential, $U_0$, it becomes expressed in terms of the
static values of the scale factor, $a_0$, and the JBD field,
$\Phi_0$. It results to be
\begin{equation}
\label{vel} \dot{\Psi}^2_0=2\,\frac{\Phi_0}{a_0^2}\,.
\end{equation}

Assuming $(3+2w)>0$, the following stability conditions were
obtained

\begin{equation} \label{cond1}{\textstyle
0<a_0^2\Phi_0\,V_0''<\frac{3}{2}\,,}
\end{equation}
and
\begin{equation}\label{cond2}{\textstyle
-\frac{3}{2}<w<-\frac{1}{4}\left[\sqrt{9-6a_0^2\Phi_0V_0''}+(3+a_0^2\Phi_0V_0'')\right].}
\end{equation}

In relation to the conditions (\ref{cond1}) and (\ref{cond2}) let
us mention that the first inequality imposes a condition on the
JBD potential, specifically for its first and second derivatives:
$0< V_0'' <V_0'/(2\Phi_0)$. The second inequality restricts the
values of the JBD parameter. Notice that this inequality imposes
that $w<0$. JBD models with negative values of $w$ have been
considered in the context of late acceleration expansion of the
universe \cite{Bertolami:1999dp, Banerjee:2000mj}, but also appear
in low energy limits of string theory \cite{Green:1987sp}.
On the other hand, as was mentioned above, we choose the JBD
potential, $V(\Phi)$, in such a way that $\Phi$ will be stabilized
at a constant value, namely $\Phi_f= m_p^2/8 \pi$.

Thus, from Eqs.~(\ref{cond1}) and (\ref{cond2}) we can conclude
that for a universe dominated by a scalar field it is possible to
obtain a static solution, stable under homogenous and isotropic
perturbation.


\section{Anisotropic perturbations}\label{secanis}

If we are interested in studying  the stability of the static
universe an important point, showed in Ref.~\cite{Barrow:2003ni},
is that the crucial destabilizing modes are not only the conformal
perturbations considered in the previous section. Anisotropic
perturbations could be even more important.
For example, in the case of the ES universe it is known
 that the static solution is
neutrally stable to inhomogeneous scalar perturbations with high
enough sound speed and to vector and tensor isotropic
perturbations\cite{Gibbons:1987jt, Barrow:2003ni}. However, this
analysis does not cover spatially homogeneous, but anisotropic
modes. It turns out that there are various unstable spatially
homogeneous anisotropic modes \cite{Barrow:2003ni}.
%
%
This suggest that anisotropic perturbations  could be the crucial
destabilizing modes.

In this section we proceed to study the stability of the static
solution found in the previous section against theses anisotropic
perturbations modes. In particular, we consider the general case
of spatially homogeneous Bianchi type IX perturbations modes.

In this context, the JBD static universe is a particular exact
solution of the Bianchi type IX, or Mixmaster universe, containing
a perfect fluid. The Mixmaster is a spatially homogeneous closed
(compact space sections) universe of the most general type. It
contains the closed isotropic Friedmann universes as particular
cases when a fluid is present. Physically, the Mixmaster universe
arises from the addition of expansion anisotropy and 3-curvature
anisotropy to the Friedmann universe.

The diagonal type IX universe has three expansion scale factors,
i.e., $a(t),\, b(t)$ and $c(t)$, and the diagonal Bianchi IX
metric is expressed by
 \be \ds
d{s}^{2}\,=\, d{t}^{2}\,-\, \eta_{\alpha \beta}(t)\,w^\alpha\,
w^\beta\,, \label{met} \en
where

\begin{equation}
\eta_{\alpha \beta}(t) = \left( \begin{array}{ccc} a^2(t) &0&0\\
0& b^2(t) &0 \\ 0& 0& c^2(t) \end{array}\right)\,,
\end{equation}

\noindent and the $w^\alpha$ are differential 1-forms invariant
under $SO(3)$ transformation.

In the following we  will consider a universe dominated by a
general perfect fluid whose equation of state is $P=(\gamma
-1)\rho$. We will assume that $\Phi$ and $\rho$ are function of
the time $t$, only. Then, by using the metric, Eq.~(\ref{met}), in
the action~(\ref{ac1}), we obtain the following set of equations
for the non-null components. The (0,0) component becomes

\begin{eqnarray}\label{00}
\frac{1}{2a^2}+\frac{1}{2b^2}+\frac{1}{2c^2} -
\frac{a^2}{4b^2c^2}- \frac{b^2}{4c^2a^2}- \frac{c^2}{4a^2b^2} +
\frac{\dot{a}\,\dot{b}}{a\, b}+ \frac{\dot{a}\,\dot{c}}{a\, c}+
\frac{\dot{b}\,\dot{c}}{b\, c} =\nonumber \\ \\\frac{\rho}{\Phi} -
\left(\frac{\dot{a}}{a}+\frac{\dot{b}}{b}+\frac{\dot{c}}{c}\right)\frac{\dot{\Phi}}{\Phi}
+ \frac{w}{2}\left(\frac{\dot{\Phi}}{\Phi}\right)^2 +
\frac{V(\Phi)}{\Phi}\,. \nonumber
\end{eqnarray}
The (1,1) component is given by

\begin{eqnarray}\label{11}
- \frac{1}{2a^2} + \frac{1}{2b^2} +\frac{1}{2c^2} -
\frac{3}{4}\frac{a^2}{b^2 c^2} +\frac{1}{4}\frac{b^2}{a^2 c^2}
+\frac{1}{4}\frac{c^2}{a^2 b^2} + \frac{\dot{b}\,\dot{c}}{b\,c} +
\frac{\ddot{b}}{b} + \frac{\ddot{c}}{c} = \nonumber \\ \\
- \frac{P}{\Phi} -
\left(\frac{\dot{b}}{b}+\frac{\dot{c}}{c}\right)\frac{\dot{\Phi}}{\Phi}
- \frac{\ddot{\Phi}}{\Phi} -
\frac{w}{2}\left(\frac{\dot{\Phi}}{\Phi}\right)^2 +
\frac{V(\Phi)}{\Phi}\,. \nonumber
\end{eqnarray}
The other two nonzero equations, components (2,2) and (3,3), are
just cyclic changes in the scale factors $(a, b, c)$ in
Eq.~(\ref{11}).

The equation for the JBD field, Eq. (\ref{f2equation}), becomes
given by
\begin{equation}\label{ecjbd}
\ddot{\Phi} +
\left(\frac{\dot{a}}{a}+\frac{\dot{b}}{b}+\frac{\dot{c}}{c}\right)\dot{\Phi}=\frac{(\rho
- 3P)}{3+2w} + \frac{2}{3+2w}\,\left[2V - \Phi
\,\frac{dV}{d\Phi}\right].
\end{equation}
On the other hand, the conservation of energy-momentum implies
that
\begin{equation}\label{energ}
\dot{\rho}+
\left(\frac{\dot{a}}{a}+\frac{\dot{b}}{b}+\frac{\dot{c}}{c}\right)(\rho+P)=0.
\end{equation}


The static solution discussed in the previous section correspond
to the case where $a(t)=b(t)=c(t)=\frac{1}{2}\,a_0$ and
$\Phi=\Phi_0$. Here, the constant values, $a_0$ and $\Phi_0$,
satisfy the conditions for a static solution. This was discussed
in Sect.~\ref{secti}.

In order to study the stability of this solution against
anisotropic Bianchi type IX perturbations, we take small
perturbations around the static solutions of the scale factors and
the JBD field. We set


\begin{eqnarray}
a(t) &=& \frac{a_0}{2}\,\big[1+\varepsilon_1(t)\big], \\
b(t) &=& \frac{a_0}{2}\,\big[1+\varepsilon_2(t)\big],
\end{eqnarray} and \begin{eqnarray}
c(t) &=&
\frac{a_0}{2}\,\big[1+\varepsilon_3(t)\big],\end{eqnarray}
together with the perturbation associated to the JBD field,
expressed by Eq.(\ref{s2b}). Here, the parameters $\varepsilon_i
(i=1,2,3)$, just like the parameter $\beta$, are small
perturbations. Therefore, they satisfy $\varepsilon_i\ll 1$.

For the energy density, pressure and state parameter we take

\begin{eqnarray}
\rho=\rho_0+\delta\rho(\varepsilon_1,\varepsilon_2,\varepsilon_3)\,, \label{pert1} \\
P=P_0+\delta P(\varepsilon_1,\varepsilon_2,\varepsilon_3)\,,
\label{pert3}
\end{eqnarray} and  \begin{eqnarray}
\gamma=\gamma_0+\delta\gamma(\varepsilon_1,\varepsilon_2,\varepsilon_3)
 \,,\label{pert2}
\end{eqnarray}
respectively. The specific form of $\delta\rho$, $\delta P$ and
$\delta\gamma$ depend on the kind of perfect fluid under
consideration.

Now, introducing these latter expressions into
Eqs.~(\ref{00}-\ref{energ}) and retaining only the linear terms in
the perturbation parameters, we obtain that

\begin{eqnarray}
-6\,\frac{\varepsilon_1}{a^2_0} + 2\,\frac{\varepsilon_2}{a^2_0} +
2\,\frac{\varepsilon_3}{a^2_0} + \ddot{\varepsilon}_2+
\ddot{\varepsilon}_3 =-\frac{\delta P}{\phi_0}+2\,\frac{\beta}{a^2_0} - \ddot{\beta}\,, \label{T1}\\
\nonumber \\
-6\,\frac{\varepsilon_2}{a^2_0} + 2\,\frac{\varepsilon_3}{a^2_0} +
2\,\frac{\varepsilon_1}{a^2_0} + \ddot{\varepsilon}_3+
\ddot{\varepsilon}_1 =-\frac{\delta
P}{\phi_0}+2\,\frac{\beta}{a^2_0} - \ddot{\beta}\,, \label{T2}
\\ \nonumber \\
-6\,\frac{\varepsilon_3}{a^2_0} + 2\,\frac{\varepsilon_1}{a^2_0} +
2\,\frac{\varepsilon_2}{a^2_0} + \ddot{\varepsilon}_1+
\ddot{\varepsilon}_2-\frac{\delta
P}{\phi_0}+2\,\frac{\beta}{a^2_0} - \ddot{\beta}\,, \label{T3}
\end{eqnarray}
and
\begin{equation}\label{T4}
(3+2w)\,\phi_0\,\ddot{\beta} =(4-3\gamma_0)\,\delta\rho -
3\delta\gamma\,\rho_0 + 2\,V_0'\,\phi_0\,\beta -
2\,\phi_0^2\,V_0''\,\beta \,.
\end{equation}

In the next subsections we study universes dominated by different
type of perfect fluids.

\subsection{Standard perfect fluid}

Here, we consider the case of a universe dominated by a standard
perfect fluid, where $\gamma$ is a constant. Then, Eqs.
(\ref{pert1}) and (\ref{pert3}) become

\begin{equation}
\rho=\rho_0+\delta\rho(\varepsilon_1,\varepsilon_2,\varepsilon_3)\approx\rho_0-
\gamma\,\rho_0 \,\big[\varepsilon_1+ \varepsilon_2+
\varepsilon_3\big]\,, \label{BP1} \end{equation} and
\begin{equation}
P=P_0+\delta P(\varepsilon_1,\varepsilon_2,\varepsilon_3)\approx
P_0 +  \gamma\,(1-\gamma)\,\rho_0\,\big[\varepsilon_1+
\varepsilon_2+ \varepsilon_3\big]\,, \label{BP2}
\end{equation}
respectively.
By introducing these expressions into Eqs.~(\ref{T1}-\ref{T4}),
and retaining the linear order in the parameters $\epsilon_i
(i=1,2,3)$ and $\beta$, we obtain a set of four coupled equations.

The general solution  of this set of equations may be written as

\begin{equation}\label{modos}
\left(
\begin{array}{c}
\varepsilon_1(t)\\
\varepsilon_2(t)\\
\varepsilon_3(t)\\
\beta(t)
\end{array}\right)= \left(
\begin{array}{c}
\bar{\varepsilon}_1\\
\bar{\varepsilon}_2\\
\bar{\varepsilon}_3\\
\bar{\beta}
\end{array}\right)e^{iwt}\;\;\;,\end{equation}
where $\bar{\varepsilon}_i$ and $\bar{\beta}$ are constants.
The frequencies corresponding to small oscillations are given by

\begin{eqnarray}\label{frec1}
w^2_1 &=& \frac{8}{a_0^2}\,, \\
w^2_2 &=& \frac{8}{a_0^2}\,, \label{frec2}
\end{eqnarray}
and
\begin{eqnarray}
\omega_{\pm}^2 &=& \frac{1}{a_0^2(3+2w)}\bigg[a_0^2\Phi_0\,V_0''
-6 + w\,(2 -
3\,\gamma) \\\nonumber \\
\label{pc2b} &\pm& \sqrt{\big[-6 + a_0^2\Phi_0V_0'' + 2w -
3w\,\gamma \big]^2  + 2(3+2w)\,\big(-6 +
a_0^2\Phi_0V_0''\,\big[3\gamma -2\big]\big)}\bigg].\nonumber
\end{eqnarray}

The static solution is stable if $\omega_{\pm}^2>0$. Assuming that
$(3+2w)>0$ we find that this solution is stable against
anisotropic perturbations, providing that
Eqs.~(\ref{cond21}-\ref{cond23}) are fulfilled.


The oscillation mode which belongs  to the frequency $w_1$ is
given by

\begin{equation}\label{V1}
\left(
\begin{array}{c}
\varepsilon_1(t)\\
\varepsilon_2(t)\\
\varepsilon_3(t)\\
\beta(t)
\end{array}\right)= C_1\left(
\begin{array}{c}
-1\\
0\\
1\\
0
\end{array}\right)e^{iw_1t}\;\;.\end{equation}
 Similarly, the oscillation mode corresponding to $w_2$
is:

\begin{equation}\label{V2}
\left(
\begin{array}{c}
\varepsilon_1(t)\\
\varepsilon_2(t)\\
\varepsilon_3(t)\\
\beta(t)
\end{array}\right)= C_2\left(
\begin{array}{c}
-1\\
1\\
0\\
0
\end{array}\right)e^{iw_2t}\;\;.\end{equation}

Finally, the oscillation modes corresponding to $w_\pm$ are:

\begin{equation}\label{V3}
\left(
\begin{array}{c}
\varepsilon_1(t)\\
\varepsilon_2(t)\\
\varepsilon_3(t)\\
\beta(t)
\end{array}\right)= C_\pm\left(
\begin{array}{c}
A_\pm\\
A_\pm\\
A_\pm\\
1
\end{array}\right)e^{iw_\pm t}\;\;,\end{equation}
where $C_i$ ($i$=1,2) and $C_\pm$ are arbitrary constants. On the
other hand, the constants $A_\pm$ are  given by

\begin{eqnarray}
A_{\pm} &=& \frac{1}{6(3\gamma - 4)}\bigg[a_0^2\Phi_0\,V_0'' -
w\,(2 - 3\,\gamma) \\\nonumber \\
\label{pc2b} &\mp& \sqrt{\big[-6 + a_0^2\Phi_0V_0'' + 2w -
3w\,\gamma \big]^2  + 2(3+2w)\,\big(-6 +
a_0^2\Phi_0V_0''\,\big[3\gamma -2\big]\big)}\bigg].\nonumber
\end{eqnarray}

Notice that the oscillation modes corresponding to the frequencies
$w_1$ and $w_2$ are anisotropic oscillations around the
equilibrium point. In these oscillations the JBD field remains
static at its equilibrium point, $\Phi_0$. On the other hand, the
oscillation modes, related  to the frequencies $w_\pm$ are
isotropic oscillations around the same point, but where now the
JBD field oscillate.

We  note that this stability behavior is completely different
wherewith it happens with the ES solution, where it was found that
spatially homogeneous Bianchi type IX modes destabilize the static
solution \cite{Barrow:2003ni}.


\subsection{Scalar field}

In this case, we consider a universe dominated by a scalar field.
Following a similar scheme to that of Sec.~\ref{static1}, we take
a flat matter potential, $U(\Psi)$, with a scalar field $\Psi$
rolling along its potential with a constant velocity satisfying
the conditions for a static universe. We study the stability of
this solution against anisotropic Bianchi type IX perturbation
modes.

In this case the set of Eqs. (\ref{pert1}-\ref{pert2}) becomes

\begin{eqnarray}
\rho&=&\rho_0+\delta\rho(\varepsilon_1,\varepsilon_2,\varepsilon_3)\approx\rho_0-
\gamma_0\,\rho_0 \,\big[\varepsilon_1+ \varepsilon_2+
\varepsilon_3\big]\,, \label{p1} \\
\nonumber\\
P&=&P_0+\delta P(\varepsilon_1,\varepsilon_2,\varepsilon_3)\approx
P_0 + \left(-\frac{2U_0}{\rho} +
\gamma_0\,(1-\gamma_0)\,\rho_0\right)\big[\varepsilon_1+
\varepsilon_2+ \varepsilon_3\big]\,. \label{p3}
\end{eqnarray}
and
\begin{eqnarray}
\gamma&=&\gamma_0+\delta\gamma(\varepsilon_1,\varepsilon_2,\varepsilon_3)\approx
\gamma_0- 2\frac{\gamma_0\,U_0}{\rho_0}\,\big[\varepsilon_1+
\varepsilon_2+ \varepsilon_3\big] \,,\label{p2}
\end{eqnarray}

We introduce these latter expressions into
Eqs.~(\ref{T1})-(\ref{T4}) and, just like above, we retain the
linear terms  in the parameters $\epsilon_i$ and $\beta$. In this
way, we obtain  a set of four coupled equations.
A general solution to this set of equations could be written as
Eq.~(\ref{modos}).

In this case, the frequencies for small oscillation are given by

\begin{eqnarray}\label{frec1}
w^2_1 &=& \frac{8}{a_0^2}\,, \\
w^2_2 &=& \frac{8}{a_0^2}\,, \label{frec2}
\end{eqnarray}
and
\begin{eqnarray}
\omega_{\pm}^2 &=&
\frac{1}{a_0^2(3+2w)}\bigg[a_0^2\Phi_0\,V_0''-2(3+2w)\label{pc2}
\\
\nonumber \\ &\pm&
\sqrt{[a_0^2\Phi_0V_0'']^2+4a_0^2\Phi_0V_0''(3+2w)+8w(3+2w)}\,\bigg].\nonumber
\end{eqnarray}

If $\omega_{\pm}^2>0$ the static solution is stable. Assuming that
$(3+2w)>0$,  we find that this solution is stable against
anisotropic perturbations, if the ranges expressed by expressions
(\ref{cond1}, \ref{cond2}) are fulfilled.
Notice that these constrains are the same constrains which were
found previously in Ref.~\cite{delCampo:2007mp}, where the
stability of this static solution against homogeneous and
isotropic perturbations was studied.

The oscillation modes corresponding to these perturbations share
similar properties than  that the ones discussed in the previous
section. In particular, they could be expressed by the same
expressions, Eqs.~(\ref{V1}, \ref{V2}, \ref{V3}), but where now
$w_1$, $w_2$ and $w_\pm$ are given by  Eqs.~(\ref{frec1},
\ref{frec2}, \ref{pc2}) respectively, and $A_\pm$ is given by

\begin{eqnarray}
A_{\pm} = \frac{1}{12}\bigg[a_0^2\Phi_0\,V_0'' + 4w \mp
\sqrt{[a_0^2\Phi_0V_0'']^2+4a_0^2\Phi_0V_0''(3+2w)+8w(3+2w)}
\bigg].\label{pc2b}
\end{eqnarray}

This modification of the stability behavior has important
consequences for the emergent universe scenario, since it
ameliorates the fine-tuning that arises from the fact that the ES
model is an unstable saddle in GR.


\section{Polynomial JBD Potential}\label{secdin}

As a particular but interesting example, we consider the case
where the JBD potential is a polynomial in the scalar field
$\Phi$.

\begin{eqnarray}\label{JBDpot}
V(\Phi) &=& C_0 + C_1\,\Phi + C_2\,\Phi^2\,,
\end{eqnarray}
where $C_0$, $C_1$ and $C_2$ are constants.
Also, we consider a homogenous and isotropic closed universe
described by a Friedmann-Robertson-Walker metric Eq.~(\ref{met1}).
As a matter content we take a standard perfect fluid.

It is interesting to notice that under these consideration and
following  the scheme of Refs.~\cite{Kolitch:1994qa,
Santos:1996jc} we can rewrite the field equations of this model,
Eqs. (\ref{key02}-\ref{key_01}), as an autonomous system.
In order to do so,  we first rewrite Eqs.~(\ref{key02},
\ref{key_01}) together with the conservation of energy equation,
by means of the conformal time

\[\eta =\int \frac{dt}{a(t)}.\]

\noindent Thus, we obtain that

\begin{eqnarray}
\left(\frac{a'}{a} + \frac{\Phi'}{2\Phi} \right)^2 + 1 &=&
\frac{\rho\,a^2}{3\,\Phi}+ \frac{2w +
3}{12}\left(\frac{\Phi'}{\Phi}\right)^2 +
\frac{C_0\,a^2}{3\,\Phi}+
\frac{V_1\,a^2}{3\,\Phi}+\frac{V_2\,a^2}{3\,\Phi}\label{DS1},
\end{eqnarray}

\begin{equation}
\frac{\Phi''}{a^2}+
2\Phi'\frac{a'}{a^3}=\frac{(4-3\gamma)}{(2w+3)}\rho +\,
\frac{2}{2w+3}\left[2V_0 + V_1\right]\;, \label{DS2}
\end{equation}
and the conservation of energy-momentum becomes
\begin{equation}
\rho'+3\,\frac{a'}{a}\,\gamma\,\rho=0 \,,
\end{equation}

\noindent where, we have used the following definitions

\begin{equation}
V_1 = C_1\,\Phi \,,\end{equation}and \begin{equation} V_2 = C_2
\,\Phi^2\,.
\end{equation}

Following Refs.~\cite{Kolitch:1994qa, Santos:1996jc} we introduce
the set of variables

\begin{eqnarray}\label{X}
X &=& \sqrt{\frac{2w + 3}{12}}\,\frac{\Phi'}{\Phi}= A
\,\frac{\Phi'}{\Phi}\,,
\\
\label{Y} Y &=& \frac{a'}{a} + \frac{\Phi'}{2\Phi}\,,
\\
Z_0 &=& \frac{C_0\,a^2}{\Phi}\,,
\\
Z_1 &=& \frac{V_1\,a^2}{\Phi}\,,
\end{eqnarray}
and
\begin{equation}
Z_2 = \frac{V_2\,a^2}{\Phi}\,.
\end{equation}

\noindent Now, we rewrite Eqs.~(\ref{DS1}) and (\ref{DS2}),
together with the energy-momentum conservation in these variables
as follows

\begin{eqnarray}\label{ligazon1}
Y^2 + 1 = \frac{\rho \,a^2}{3\Phi} + X^2 + \frac{Z_0}{3}+
\frac{Z_1}{3}+ \frac{Z_2}{3}\,, \end{eqnarray}\\
and
\begin{eqnarray}
X' = -2X\,Y + \left(1-\frac{3}{4}\gamma\right)\!\frac{\rho
\,a^2}{3A\,\Phi}+ \frac{Z_0}{3A}+ \frac{Z_1}{6A}\,.
\end{eqnarray}

Differentiating Eq.~(\ref{ligazon1}) and from the equation for the
$X$ variable, together with the energy-momentum conservation, we
obtain that

\begin{eqnarray}\label{dinamicals1}
X' &=& -2X\,Y + \left(1-\frac{3}{4}\gamma\right)
\left[\frac{Y^2+ 1 -X^2}{A}\right]+\frac{\gamma}{4}\frac{Z_0}{A} \nonumber \\
\nonumber \\
&&+ \left(\frac{\gamma}{4}-\frac{1}{6}\right)\frac{Z_1}{A}+
\left(\frac{3}{4}\gamma- 1\right)\frac{Z_2}{3A}\,, \\
\nonumber\\
Y' &=& -2X^2 + \left(1-\frac{3}{2}\gamma\right)\left[Y^2+ 1
-X^2\right] +
\frac{\gamma}{2}\left(Z_0 + Z_1 + Z_2\right)\,,\label{dinamicals2}\\
\nonumber \\
Z'_0 &=& 2Z_0\left[-\frac{X}{A}+Y\right],\label{dinamicals3}\\
\nonumber \\
Z'_1 &=& 2Z_1\left[-\frac{X}{2A}+Y\right],\label{dinamicals4}
\end{eqnarray} and \begin{equation}
Z'_2 = 2Z_2\,Y\,\label{dinamicals5}.
\end{equation}

Requiring that $\rho >0$, we get from Eq.~(\ref{ligazon1}) that
\begin{equation}\label{ligazon2}
Y^2 - X^2 - \frac{Z_0}{3}- \frac{Z_1}{3}- \frac{Z_2}{3}+ 1 \geq0
\,.
\end{equation}
In the set of  equations (\ref{dinamicals1})-(\ref{dinamicals5})
we look for critical points. In particular, we are interested in
critical points related to static universes which were discussed
in Sect.~\ref{static1}. Thus, from
Eqs.~(\ref{dinamicals1})-(\ref{dinamicals5}) together with
expression (\ref{X}) and (\ref{Y}), the critical points correspond
to $X = Y = 0$, $Z_0 =\bar{Z}_0$, $Z_1 = \bar{Z}_1$ and $Z_2
=\bar{Z}_2$, where
\begin{eqnarray}
\bar{Z}_0 &=& \frac{3}{2} - \frac{2}{\gamma} -
\frac{\bar{Z}_1}{2},
\end{eqnarray}
and  \hspace{-6cm}{\begin{equation}
 \bar{Z}_2 = \frac{3}{2}  - \frac{\bar{Z}_1}{2}.
\end{equation}}
\noindent Then,  we have a set of critical points which represents
different static universes. They depend on the arbitrary value of
$\bar{Z}_1$.  Actually, the possibility of obtaining stable or
instable critical points depends on the value of $\bar{Z_1}$. In
the following, we will give a range for the parameter  $\bar{Z_1}$
where the corresponding solutions are stable (see Eqs.(\ref{84})
and (\ref{d2}))

\begin{figure}[h]
\begin{center}
\includegraphics[width=4.4in,angle=0,clip=true]{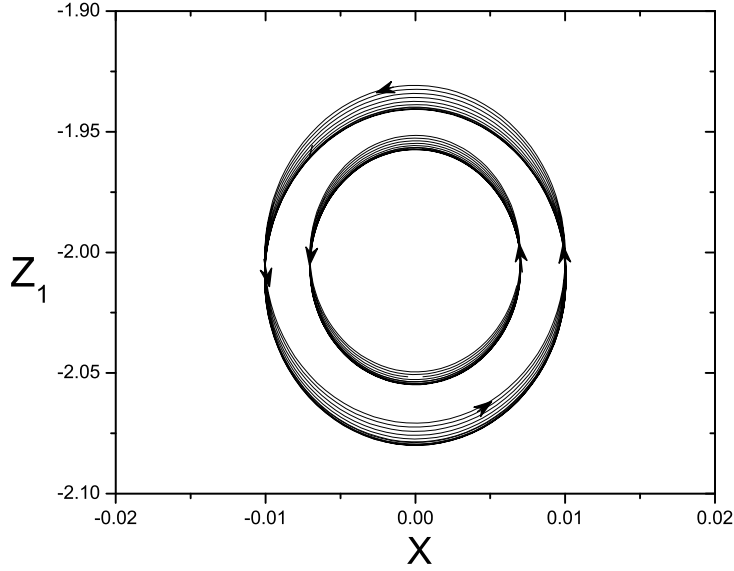}
\caption{Plot showing the evolution of two numerical solutions for
a universe dominated by dust.} \label{Fdust}
\end{center}
\end{figure}

\begin{figure}[h]
\begin{center}
\includegraphics[width=4.4in,angle=0,clip=true]{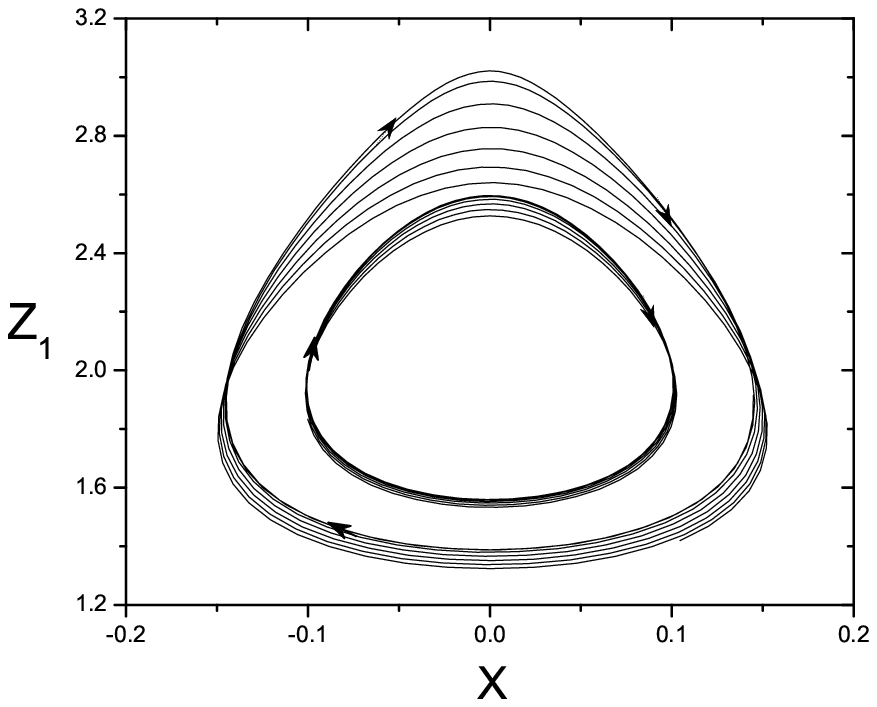}
\caption{Plot showing the evolution of two numerical solutions for
a universe dominated by a scalar field.} \label{Fscalar}
\end{center}
\end{figure}

In order to study the nature of these critical points we linearize
the set of equations (\ref{dinamicals1}-\ref{dinamicals5}) near
the critical points. From the study of the eigenvalues of the
system we found that the critical points could be centers or
saddles points, depending on the values of the parameters of the
model ($\gamma$ and $A$) and on the value of $\bar{Z}_1$.
Stable static solutions correspond to a center, and this imposes
the following conditions for the parameters $A$ and $\gamma$, and
for the value of $\bar{Z}_1$.

\begin{eqnarray}
\frac{2}{3}< &\gamma& < \frac{4}{3} \,,\label{d1}\\
0 < &A& < \frac{4-3\gamma}{6\gamma -4}\,,\end{eqnarray} and
\begin{eqnarray} \frac{9\gamma - 12}{3\gamma -2} < &\bar{Z}_1& <
-6(1 + 2A(2 + A)) + \frac{9}{2}\,(1 + 2A)^2\, \gamma \,,\label{84}
\end{eqnarray}

or

\begin{eqnarray}
\frac{4}{3}< &\gamma& <\infty\,,\\
0 < &A& < \frac{3\gamma -4}{6\gamma -4}\,,\end{eqnarray} and
\begin{eqnarray} \frac{9\gamma - 12}{3\gamma -2} < &\bar{Z}_1& < -6(1 -2A(2
- A)) + \frac{9}{2}\,(1 - 2A)^2\, \gamma \,.\label{d2}
\end{eqnarray}

These conditions are in agrement with the general stability
conditions that were found previously in Sect.\ref{static1} (see
Eqs.~(\ref{cond21}-\ref{cond23})).


In Fig.~\ref{Fdust} it is shown a projection of the axis $X$ and
$Z_1$. This represents the evolution of two numerical solutions
for a universe dominated by dust. In order to satisfy the
requirements of stability we have taken the values $A=0.008$ and
$\bar{Z}_1 = -2$. Here, the critical point, which in this graph
corresponds to the point $X=0$ and $Z_1=\bar{Z}_1 = -2$,
represents a center.

In Fig.~\ref{Fscalar} it is shown a projection of the axis $X$ and
$Z_1$ of two numerical solutions for the case where the universe
is dominated by a scalar field moving in a null scalar potential.
In order to satisfy the requirements of stability  we take
$A=0.008$ and $\bar{Z}_1 = 2$. As we expect, the critical point,
which in this graph correspond to the point $X=0$ and
$Z_1=\bar{Z}_1= 2$, it is a center.

\section{Conclusions}
\label{conc}

In this paper, we have studied the stability of the JBD static
universe model. This is motivated by the possibility that the
universe might have started out in an asymptotically JBD static
state, in the context of the so called emergent universe models.

We extent our previous results on stability of JBD static universe
by considering  spatially homogeneous Bianchi type IX anisotropic
perturbation modes and by including  more general perfect fluid.
Contrary to GR we have found that the JBD static universe
dominated by a standard perfect fluid could be stable against
isotropic and anisotropic perturbations for some sort of perfect
fluids, for example for dust or scalar field (inflaton).
 This modification of the stability behavior has important
consequences for the emergent universe scenario, since it
ameliorates the fine-tuning that arises from the fact that the ES
model is an unstable saddle in GR and prevent that small
fluctuations, such as quantum fluctuations, will inevitably arise,
forcing the universe away from its static state, thereby aborting
the emergent universe scenario.

In particular we found that for a standard perfect fluid with a
polytropic state equation satisfying $\gamma > 2/3$ it is possible
to find a static solution which is stable against isotropic and
anisotropic perturbations, with the only exception of radiation
($\gamma = 4/3$).
%
This implies that for a universe dominated by dust ($\gamma =1$),
for example, we could find a solution where the universe is static
and stable.
 The instability of the static universe dominated by
radiation, although disturbing, seems  not to be a problem,  since
in a pre-inflationary cosmological model, it might be possible
that radiation be an element which is not dominant  at all.

Also, we found that the static JBD universe described in
Ref.~\cite{delCampo:2007mp}, which correspond to a universe
dominated by a scalar field moving in a flat potential, is stable
against isotropic and anisotropic perturbations when the JBD
potential and the JBD parameter satisfy  a set of general
conditions discussed in Sect.~\ref{secanis}.



Finally, we focus on a particular example of a JBD potential,
which is a polynomial in the JBD field. This kind of JBD potential
allow us to use a dynamical system approach for studying the
stability of the JBD static universe. In this respect, we have
found that the JBD static universe solutions are center
equilibrium points.
We obtained numerical solutions for a universe dominated by
standard perfect fluids and dominated by a scalar field. We have
considered the cases where the universe starts from an initial
state close to the equilibrium point. The numerical solutions
showed a behavior just like the expected if the equilibrium points
are centers.

We should  stress that in this work we have studied the stability
of the Jordan-Brans-Dicke  static universe against spatially
homogeneous isotropic and anisotropic perturbations, see Refs.
\cite{Gibbons:1987jt, Barrow:2003ni}. Of course, also it is
possible to study  the stability of the JBD static universe
against  spatially inhomogeneous perturbations (scalar, vector and
tensor perturbations). The situation for the ES solution, becomes
 neutrally stable against inhomogeneous scalar perturbations (with
high enough sound speed),  vector and tensor isotropic
perturbations \cite{Gibbons:1987jt, Barrow:2003ni}.
 We expect that in the JBD case  these inhomogeneous
perturbations do not lead to additional instabilities in the same
way that happens with the ES case. We intend to return to this
point in the near future by working an approach analogous to that
followed in Refs. \cite{Bruni:1992dg,Dunsby:1991xk,
Bruni:1991kb,Dunsby:1998hd}.

\begin{acknowledgments}
This work was supported by the COMISION NACIONAL DE CIENCIAS Y
TECNOLOGIA through FONDECYT Grant N$^{0}$s. 1070306 (SdC) and
1090613 (RH and SdC) and also was partially supported by PUCV
Grant N$^0$ 123.787/2007 (SdC) and N$^0$ 123.703/2009 (RH). PL was
supported by Direcci\'on de Investigaci\'on de la Universidad del
B\'{\i}o-B\'{\i}o through Grant N$^0$ 096207 1/R. PL wishes to
thank the warm hospitality extended to him during his visits to
Institute of Cosmology and Gravitation, University of Portsmouth
were part of this work was done.

\end{acknowledgments}


\end{document}